\pdfoutput=1

\documentclass[sigconf]{acmart}
\AtBeginDocument{%
  }

 \copyrightyear{2026}
 \acmYear{2026}
 \setcopyright{cc}
 \setcctype{by}
 \acmConference[MM '26]{Proceedings of the 34th ACM International Conference on Multimedia}{November 10--14, 2026}{Rio de Janeiro, Brazil}
 \acmBooktitle{Proceedings of the 34th ACM International Conference on Multimedia (MM '26), November 10--14, 2026, Rio de Janeiro, Brazil}
 \acmDOI{10.1145/3767308.3832542}
 \acmISBN{979-8-4007-2213-4/2026/11}

\acmSubmissionID{111}

\usepackage{xspace}
\usepackage{multirow}
\usepackage{booktabs}
\usepackage{graphicx}
\usepackage{subcaption}
\usepackage{cleveref}
\usepackage{enumitem}

\begin{document}

\makeatletter
\DeclareRobustCommand\onedot{\futurelet\@let@token\@onedot}
\def\@onedot{\ifx\@let@token.\else.\null\fi\xspace}

\def\eg{\emph{e.g}\onedot} \def\Eg{\emph{E.g}\onedot}
\def\ie{\emph{i.e}\onedot}
\def\Ie{\emph{I.e}\onedot}
\def\cf{\emph{cf}\onedot} \def\Cf{\emph{Cf}\onedot}
\def\etc{\emph{etc}\onedot} \def\vs{\emph{vs}\onedot}
\def\wrt{w.r.t\onedot} \def\dof{d.o.f\onedot}
\def\iid{i.i.d\onedot} \def\wolog{w.l.o.g\onedot}
\def\etal{\emph{et al}\onedot}
\makeatother

\definecolor{lightmauve}{rgb}{0.86, 0.82, 1.0}
\definecolor{lightgoldenrodyellow}{rgb}{0.98, 0.98, 0.82}
\definecolor{lightapricot}{rgb}{0.99, 0.84, 0.69}
\definecolor{lightblue}{rgb}{0.55, 0.85, 0.9}
\definecolor{lightskyblue}{rgb}{0.53, 0.81, 0.98}
\definecolor{non-photoblue}{rgb}{0.64, 0.87, 0.93}
\definecolor{lightcornflowerblue}{rgb}{0.6, 0.81, 0.93}
\definecolor{lightgreen}{rgb}{0.56, 0.93, 0.56}
\definecolor{lightseagreen}{rgb}{0.13, 0.7, 0.67}
\definecolor{lightpink}{rgb}{1.0, 0.71, 0.76}
\definecolor{purered}{rgb}{1.0, 0.0, 0.0}
\definecolor{fontblue}{rgb}{0.0, 0.0, 1.0}
\definecolor{secondorange}{rgb}{0.85, 0.4, 0.0}
\definecolor{secondpurple}{rgb}{0.5, 0.0, 0.5}

\title{Beyond Distortion Robustness: Rethinking Severe Cropping as Erasure-Resilient Message Embedding}

\author{Bo Pang}
\authornote{ Denotes equal contribution.}
\email{bo_pang@hrbeu.edu.cn}
\orcid{0009-0006-6354-6766}   

\affiliation{%
  \institution{Harbin Engineering University}
  \city{Harbin}
  \state{Heilongjiang}
  \country{China}
}

\author{Weibin Kong}
\authornotemark[1]
\email{3360731163@hrbeu.edu.cn}
\orcid{0009-0006-5269-2207}  

\affiliation{%
  \institution{Harbin Engineering University}
  \city{Harbin}
  \state{Heilongjiang}
  \country{China}
}

\author{Juntu Dong}
\authornotemark[1]
\email{1930312281@hrbeu.edu.cn}
\orcid{0009-0009-0581-4242}  

\affiliation{%
  \institution{Harbin Engineering University}
  \city{Harbin}
  \state{Heilongjiang}
  \country{China}
}

\author{Minghan Li}
\email{minghan@hrbeu.edu.cn}
\orcid{0009-0004-9213-9217}  

\affiliation{%
  \institution{Harbin Engineering University}
  \city{Harbin}
  \state{Heilongjiang}
  \country{China}
}

\author{Zhongping Zhang}
\authornote{Corresponding author.}
\email{zhongping@hrbeu.edu.cn}
\orcid{0000-0002-6274-190X}  

\affiliation{%
  \institution{Harbin Engineering University}
  \city{Harbin}
  \state{Heilongjiang}
  \country{China}
}

% \author{Juntu Dong}
% \affiliation{%
%   \institution{Harbin Engineering University}
%   \city{Harbin}
%   \email{1930312281@hrbeu.edu.com}

% \author{Weibin Kong}
% \affiliation{%
%   \institution{Harbin Engineering University}
%   \city{Harbin}
%   \email{weibin_kong_2026@163.com}

\begin{abstract}
Robust message embedding in images is important for multimedia security applications such as copyright protection and content tracing. Existing methods are largely developed under a \emph{distortion robustness} paradigm, where the embedded signal remains spatially present but is degraded by noise, blur, or compression. Severe cropping poses a fundamentally different challenge because it removes part of the carrier itself, causing partial payload disappearance rather than mere signal corruption. In this paper, we revisit robust message embedding from an \emph{erasure-resilience} perspective and present CREST, a proof-of-concept framework for severe-cropping-robust embedding. CREST combines coding-theoretic redundancy with neural embedding and recovery by expanding a compact QR message into a redundant spatial payload via LT fountain coding and coupling it with cropping-aware embedding and fragment recovery. Experiments on COCO, DIV2K, and VOC2012 show that CREST improves recovery under severe cropping while maintaining competitive visual quality. Under mixed distortions with an area retention ratio of 0.7, CREST improves TRA from 18.52\% to 68.45\% and reduces EMR from 13.88\% to 4.21\% over the strongest baseline. On COCO2017, CREST still achieves 48.55--65.12\% TRA when only 30--50\% of the image area is retained, whereas all compared baselines fail to recover the message\footnote{Code: https://github.com/ZPZhangLab/CREST-ACMMM2026}. These results suggest that severe cropping is better understood as an erasure problem rather than a conventional distortion problem, motivating the joint design of neural embedding and coding-based recovery.
\end{abstract}

\begin{CCSXML}
<ccs2012>
   <concept>
       <concept_id>10002978.10002979.10002984</concept_id>
       <concept_desc>Security and privacy~Information-theoretic techniques</concept_desc>
       <concept_significance>500</concept_significance>
       </concept>
   <concept>
       <concept_id>10002978.10002991.10002996</concept_id>
       <concept_desc>Security and privacy~Digital rights management</concept_desc>
       <concept_significance>300</concept_significance>
       </concept>
   <concept>
       <concept_id>10010147.10010257.10010293.10010294</concept_id>
       <concept_desc>Computing methodologies~Neural networks</concept_desc>
       <concept_significance>300</concept_significance>
       </concept>
 </ccs2012>
\end{CCSXML}

\ccsdesc[500]{Security and privacy~Information-theoretic techniques}
\ccsdesc[300]{Security and privacy~Digital rights management}
\ccsdesc[300]{Computing methodologies~Neural networks}


\keywords{robust message embedding, image steganography, multimedia security, fountain codes, normalizing flow, cropping resilient}
  %%\includegraphics[width=\textwidth]{sampleteaser}
  %%\caption{Seattle Mariners at Spring Training, 2010.}
  %%\Description{Enjoying the baseball game from the third-base
  %%seats. Ichiro Suzuki preparing to bat.}
  %%\label{fig:teaser}


\maketitle

\begin{figure}[htbp] 
  \centering         
  \includegraphics[width=0.90\linewidth]{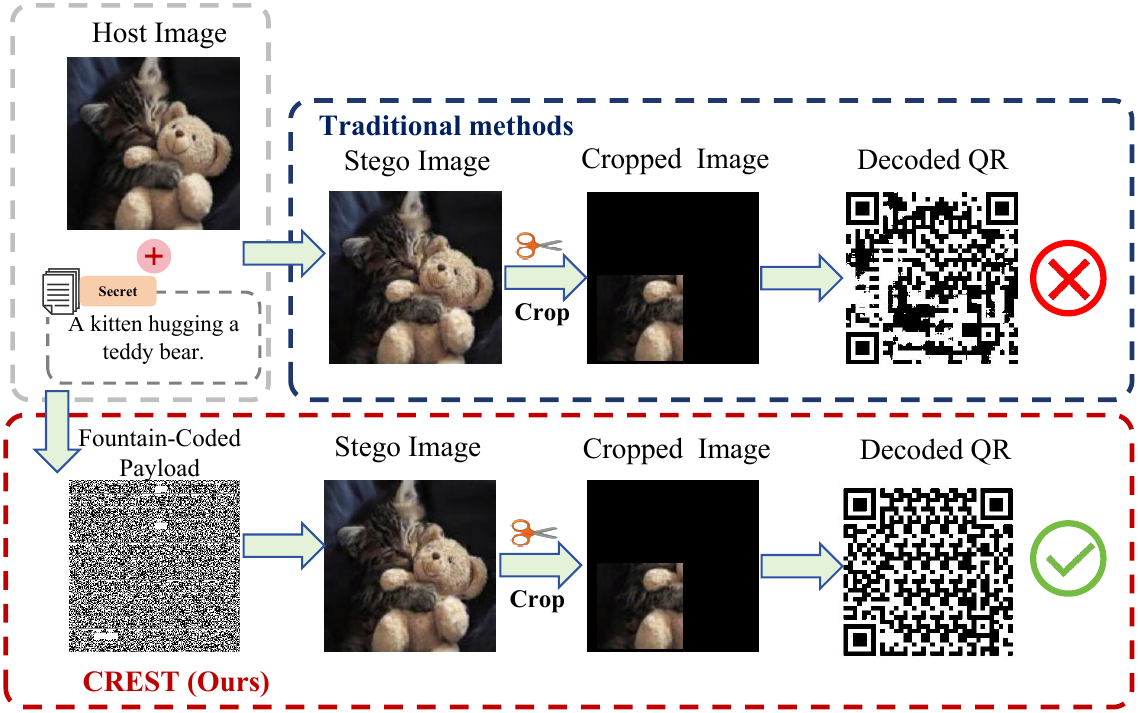}
  \caption{Under severe cropping and global distortions, conventional distortion-robust methods degrade sharply (top), whereas CREST maintains more reliable decoding through erasure-resilient message embedding (bottom).}  
  \Description{A comparison showing that conventional embedding fails to recover the QR code after severe cropping, whereas CREST successfully reconstructs a decodable QR code.}
  \label{fig:overview_comparison}
\end{figure}

\section{Introduction}
\label{sec:intro}

Robust message embedding in images is a core problem in multimedia security, with applications including copyright protection~\cite{zhu2018hidden,jing2021hinet}, content tracing~\cite{wang2024must,liang2025screenmark}, and physical-to-digital content linking~\cite{tancik2020stegastamp,fu2022chartstamp}. Recent learning-based methods have significantly improved the ability to hide machine-readable information in natural images while surviving common degradations such as JPEG compression, Gaussian noise, blur, and printing~\cite{zhu2018hidden,tancik2020stegastamp,xu2022riis,fu2022chartstamp,lan2023freq,shadmand2024stampone,yang2024provablyrobust,peng2024ldstega,qi2025crossmodal,ye2025rmsteg}. In most of these settings, however, the embedded signal is still spatially present: it may be weakened, blurred, or contaminated, but it is not physically removed. As a result, existing robust embedding pipelines are predominantly developed under a \emph{distortion robustness} paradigm~\cite{zhu2018hidden,xu2022riis,lan2023freq,shadmand2024stampone,yang2024provablyrobust,qi2025crossmodal}, as shown in the top row of \Cref{fig:overview_comparison}.

Severe local cropping does not fit this paradigm well. Unlike global distortions that perturb the image while preserving its spatial support, cropping directly removes part of the carrier and therefore destroys part of the embedded payload itself~\cite{ma2025ropass,yang2025screenshoot,liu2025postencoding}. Once a sufficiently large region is discarded, decoding can fail catastrophically because the remaining fragments no longer contain enough usable information for reliable recovery. From this viewpoint, cropping is fundamentally not just another distortion. Rather, it is a form of \emph{spatial erasure}: the key difficulty lies not only in denoising or synchronization, but in recovering a message after part of its support has disappeared~\cite{ma2025ropass,liu2025postencoding}.

This observation motivates the central premise of this paper: \emph{cropping robustness in message embedding should be revisited from an erasure-resilience perspective}. When severe cropping behaves like partial payload erasure, coding-theoretic redundancy becomes a natural design tool~\cite{byers1998digitalfountain,luby2002lt,keller2022fountain,yao2024ldgm,yao2025nestedpolar}. In principle, one can spread the message across a larger redundant spatial representation so that recovery remains possible from surviving image fragments. However, this idea also introduces a second challenge. Enlarging the payload substantially increases the amount of information that must be embedded, which raises model complexity, computational cost, and perceptual risk~\cite{fu2022chartstamp,apau2024slr,ye2025rmsteg}. Therefore, the problem is not simply how to improve robustness under cropping, but how to jointly address \emph{cropping-induced information loss} and \emph{redundancy-induced embedding overhead}.

In this paper, we instantiate this perspective through CREST, a \textbf{C}ropping-\textbf{R}esilient and \textbf{E}fficient \textbf{ST}eganographic framework, as shown in \Cref{fig:overview_comparison} (bottom row). Rather than treating cropping as a minor extension of distortion robustness, CREST serves as a proof-of-concept system for \emph{fountain-coded erasure-resilient message embedding}~\cite{byers1998digitalfountain,luby2002lt,keller2022fountain}. At the message level, the secret message is expanded into a redundant spatial payload through LT fountain coding, enabling recovery from partial observations~\cite{luby2002lt,byers1998digitalfountain,keller2022fountain,tancik2020stegastamp,ye2025rmsteg}. At the extraction level, pilot-guided localization and residual enhancement are introduced to stabilize tile parsing and improve the reliability of fragment recovery under severe cropping~\cite{tancik2020stegastamp,shadmand2024stampone,xu2022riis,cao2024channelattn}. At the training level, cropping-aware curriculum learning further strengthens robustness under progressively harsher erasure conditions~\cite{tancik2020stegastamp,fu2022chartstamp,shadmand2024stampone,ye2025rmsteg}. Finally, we incorporate a lightweight distilled attention-flow design to reduce model complexity introduced by the redundant payload.

This formulation leads to a different way of thinking about robust multimedia embedding. Under the conventional view, cropping is often handled as a special case of geometric distortion or synchronization failure~\cite{zhu2018hidden,tancik2020stegastamp,lan2023freq}. Such a treatment may remain effective when the cropped region is relatively small, but it becomes increasingly inadequate under severe cropping, where a substantial fraction of the payload may simply disappear~\cite{ma2025ropass,yang2025screenshoot,liu2025postencoding}. Under our view, cropping should instead be addressed through the joint design of \emph{neural embedding}, \emph{spatial redundancy}, and \emph{coding-based reconstruction}~\cite{byers1998digitalfountain,luby2002lt,keller2022fountain,qi2025crossmodal}. This shift is important because it changes both the objective and the system design: the goal is no longer merely to preserve a fragile hidden signal under perturbation, but to ensure recoverability even when part of the embedded support is physically absent.

Extensive experiments on COCO, DIV2K, and VOC2012 validate this perspective. The proposed framework achieves substantially stronger recovery under severe cropping while maintaining competitive visual quality and favorable efficiency. For example, under mixed distortions with an area retention ratio of 0.7, CREST improves TRA from 18.52 to 68.45 and reduces EMR from 13.88 to 4.21 compared with state-of-the-art baselines. On COCO2017, when only 30--50\% of the image area is retained, CREST still achieves a TRA around 48--65, whereas competing methods fail to recover the message. More importantly, the results suggest a broader research direction: robust message embedding under localized content loss may be better approached as a hybrid problem at the intersection of multimedia representation learning and erasure-resilient communication, rather than as a straightforward extension of distortion-robust steganography~\cite{ye2025rmsteg,qi2025crossmodal,yao2025nestedpolar}.

Our main contributions are summarized as follows:
\begin{itemize}[nosep,leftmargin=*]
    \item We argue that severe cropping in robust message embedding is better understood as a \emph{spatial erasure} problem rather than a conventional distortion problem, and we introduce an erasure-resilience perspective for this setting.
    \item We present CREST, a proof-of-concept robust embedding framework that combines LT-coded redundant payload construction, pilot-guided localization, enhanced residual recovery, and cropping-aware training for severe-cropping scenarios.
    \item We show that the redundancy required for erasure resilience can be coupled with a lightweight distilled attention-flow design to achieve a more favorable robustness--efficiency trade-off.
    \item Experiments on multiple datasets demonstrate that CREST substantially improves recovery under severe cropping while maintaining competitive visual quality and efficiency.
\end{itemize}

\begin{figure*}[t]
  \centering
  \includegraphics[width=0.7\linewidth]{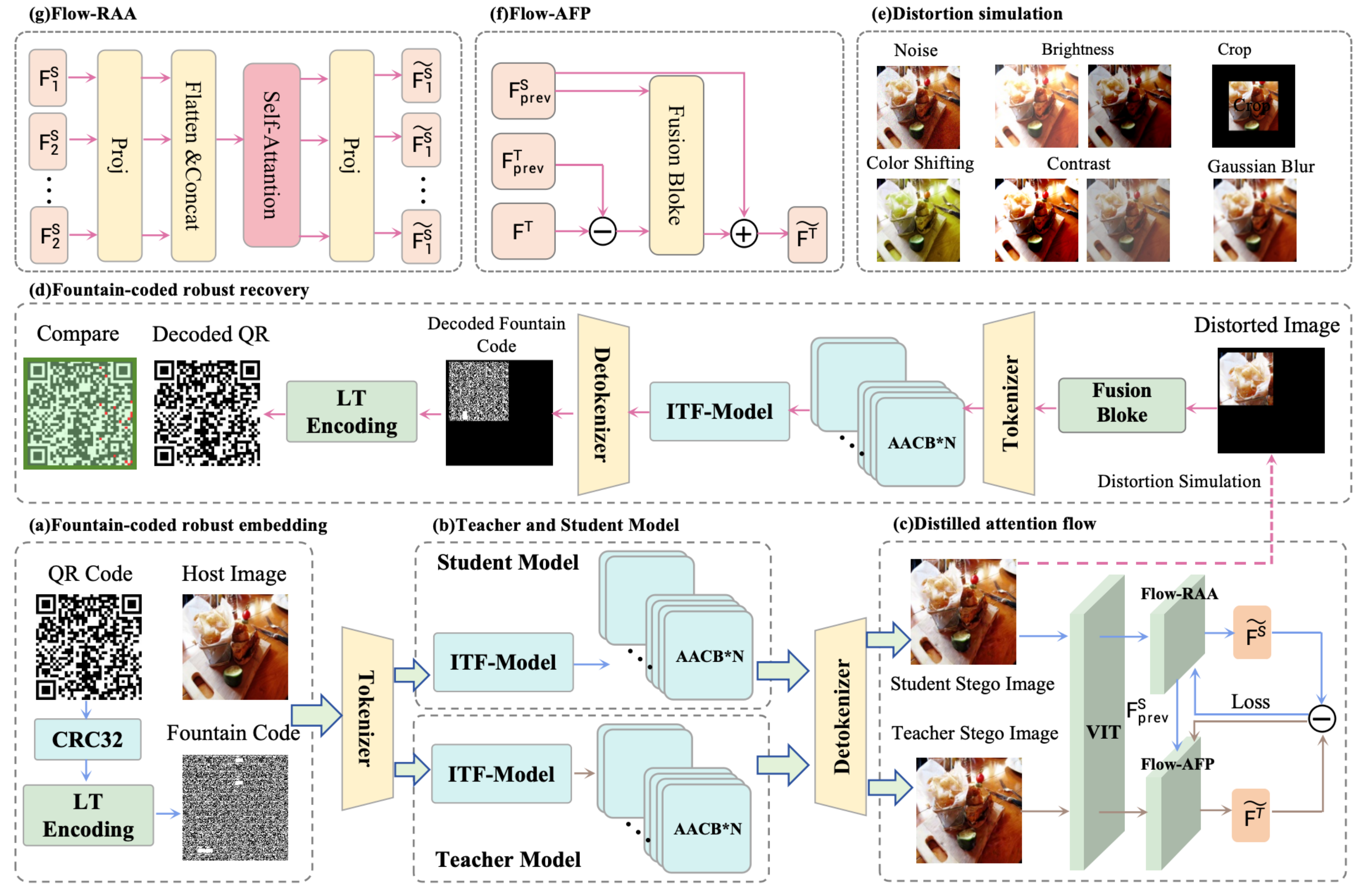}
  \caption{Overview of CREST. (a) A CRC32-augmented QR message is LT-encoded into a redundant fountain-code payload. (b) Teacher and student ITF-AACB branches embed the payload into the host image. (c) Training-time distillation aligns their ViT features using Flow-RAA and Flow-AFP. (d) Inverse embedding, pilot-guided parsing, LT decoding, and final CRC32 verification recover the QR message. (e) Training applies noise, color/brightness/contrast variations, blur, JPEG compression, and cropping. (f--g) Flow-AFP and Flow-RAA construct the teacher target and aligned student features, respectively.}
  \Description{An overview of CREST, including fountain-coded payload construction, teacher--student invertible embedding, feature distillation, distortion simulation, and LT-based message recovery.}
  \label{fig:framework}
\end{figure*}

\section{Related Work}
\label{sec:related_work}

\noindent \textbf{Robust Image Steganography}.
Recent robust image steganography has largely evolved toward deep end-to-end and invertible embedding frameworks~\cite{zhu2018hidden,lu2021isn,jing2021hinet}. 
Subsequent studies improve robustness through physical-world embedding~\cite{tancik2020stegastamp}, application-oriented embedding~\cite{fu2022chartstamp}, invertible robust hiding~\cite{xu2022riis}, frequency-aware design~\cite{lan2023freq}, printer-proof optimization~\cite{shadmand2024stampone}, and generative or attention-flow-based models~\cite{peng2024ldstega,ye2025rmsteg}. 
More recent work further considers coding-aware or provable robustness~\cite{yang2024provablyrobust,qi2025crossmodal}. Related multimedia learning tasks also investigate image manipulation, fine-grained signal analysis, and efficient representation~\cite{zhang2024text,zhang2024machine,zhang2024movie}. Despite these advances, most robust embedding methods assume that the hidden signal remains spatially present but is degraded by noise, blur, compression, or other distortions~\cite{ye2025rmsteg}. This assumption becomes less suitable when severe cropping physically removes part of the embedded support.

\noindent \textbf{Cropping Robustness and Erasure-Resilient Recovery.}
Cropping-robust watermarking and message embedding have been studied under screen-shooting, partial-capture, and geometric-attack scenarios~\cite{fang2022pimog,fang2023fin,wang2024must,liang2025screenmark,liu2025postencoding,ma2025ropass,cao2024channelattn}. Most existing approaches primarily address synchronization or localization, whereas severe cropping additionally raises the question of whether sufficient information survives for decoding. This connects the problem to erasure-resilient communication, where digital fountain and LT codes recover messages from sufficient subsets of encoded symbols~\cite{byers1998digitalfountain,luby2002lt}. Coding-based redundancy has also been explored in watermarking and steganography~\cite{korus2014fountain,keller2022fountain,yao2024ldgm,yao2025nestedpolar,qi2025crossmodal,yang2024provablyrobust}, but remains only weakly integrated with modern neural embedding under severe spatial loss.

\noindent \textbf{Efficiency in Flow-Based Embedding.}
Efficiency in flow-based embedding differs from generic model compression because the same structured mapping must support both forward embedding and stable inverse recovery~\cite{gomez2017revnet,jacobsen2018irevnet,lu2021isn,jing2021hinet,xu2022riis,ye2025rmsteg}. 
Aggressive compression may therefore impair inverse stability, a known issue in invertible networks~\cite{behrmann2021exploding}. 
Existing efficiency-oriented approaches typically preserve the invertible structure through distillation or efficient flow designs~\cite{oord2018parallelwavenet,walton2025distillnf,hoogeboom2019idf,gritsenko2021idfpp,wang2022iodf}. 
In our setting, efficiency is meaningful when computational savings do not compromise message recoverability under distortion and cropping.

\section{Method}
\label{sec:method}

\subsection{Overview}
We propose \textbf{CREST}, an end-to-end robust message embedding framework for severe-cropping scenarios. The central idea is to address cropping-induced information loss from an \emph{erasure-resilience} perspective while explicitly controlling the embedding overhead introduced by redundancy. Unlike conventional robust embedding pipelines that mainly target degradations (\eg, noise, blur, or compression), CREST further considers the case where part of the embedded spatial support may be physically removed.

As illustrated in \Cref{fig:framework}, CREST couples robustness at the message, spatial, and neural embedding levels. At the message level, the secret payload is transformed into redundant coded symbols through LT fountain coding to support recovery from partial observations (\Cref{sec:method_fountain_coded}). At the spatial level, each coded symbol is mapped to a self-descriptive tile with pilot patterns and metadata, enabling localization, parsing, and fragment selection after cropping (\Cref{sec:method_spatial_tiling}). At the neural embedding level, the tiled payload is hidden into the host image through an attention-flow-based backbone, together with a distilled embedding strategy and a residual enhancement module that improve extraction stability and deployment efficiency (\Cref{sec:method_distilled_embedding,sec:method_enhancer}). The surviving valid tiles are then assembled into a linear system over $GF(2)$ for message reconstruction (\Cref{sec:method_gf2}). From this viewpoint, CREST separates the problem into two coupled stages: the survival of recoverable coded fragments under cropping, and exact message reconstruction from those fragments. The resulting rank-based recoverability perspective also motivates our progressive curriculum design (\Cref{sec:method_rank_analysis}).

\subsection{Fountain-Coded Payload Construction}
\label{sec:method_fountain_coded}

\noindent \textbf{Payload Preparation.} Given a secret message $M \in \{0,1\}^{1369}$, represented here as a $37 \times 37$ binary payload grid consistent with a Version-5-style QR layout, we first append a 32-bit Cyclic Redundancy Check (CRC) for end-to-end integrity verification:
\begin{equation}
D = M \| \text{CRC32}(M),
\end{equation}
where $\|$ denotes concatenation. The augmented bitstream $D$ is then zero-padded to a multiple of the predefined block length $L = 192$ bits and partitioned into $K$ source blocks. In our implementation,
\begin{equation}
K = \left\lceil \frac{|D| + p}{L} \right\rceil = 8,
\end{equation}
where $p$ is the padding length. Each source block is denoted as $B_k \in \{0,1\}^{192}$.

\noindent \textbf{LT Encoding with Systematic Redundancy.}
Although Luby Transform (LT) codes\cite{4623778} are theoretically rateless, the finite spatial resolution of the visual carrier constrains the number of embedding symbols that can be instantiated in practice. We therefore truncate the fountain stream to a fixed grid of $N = 196$ symbols, corresponding to a redundancy factor of $24.5\times$ ($196/8$). Each encoded symbol $S_i$ is computed as the XOR sum of a pseudo-randomly selected subset of source blocks:
\begin{equation}
S_i = \bigoplus_{j \in \mathcal{S}_i} B_j,
\end{equation}
where the degree $d_i = |\mathcal{S}_i|$ is sampled from a modified robust soliton distribution. For the ultra-short block regime considered here ($K=8$), standard soliton distributions often do not accumulate full rank efficiently under heavy erasures. We therefore empirically optimized the degree distribution via Monte Carlo simulation to maximize decoding probability in this setting: $P(1)=0.10,\quad P(2)=0.35,\quad P(3)=0.30,\quad P(4)=0.15,\quad P(5)=0.07$, and $P(d \sim \text{Uniform}(1,K))=0.03$. 

To accelerate the waterfall phase of LT decoding under severe cropping, we further impose \textbf{systematic redundancy}: the first $4K = 32$ symbols are forced to be unmodified copies of the original source blocks ($d=1$), increasing the probability that the base rank is preserved in the surviving spatial fragments.

\subsection{Spatial Tiling and Aperiodic Pilot Alignment}
\label{sec:method_spatial_tiling}

\noindent \textbf{Tile Bit Layout and Symbol Permutation.} Each symbol is visually mapped to a $16 \times 16$ pixel tile, allocating 256 bits. Rows 0--1 are reserved for alignment pilots (32 bits). Rows 2--15 encode the symbol metadata---Seed (16 bits), Degree (8 bits), and 8 auxiliary filler bits---followed by the 192-bit payload $S_i$. This self-descriptive layout ensures that any surviving tile can be parsed independently. To prevent a localized crop from destroying all systematic blocks, we apply a globally fixed permutation $\pi = \text{PRNG}_{\text{seed}}(\{0,\dots,195\})$ to uniformly scatter the symbols across the $14 \times 14$ image grid.

\noindent \textbf{Aperiodic Pilot Pattern.} Geometric distortions and cropping destroy the global coordinate system, introducing phase ambiguity during tile extraction. We therefore place an \textbf{aperiodic pilot sequence} in the top two rows of every tile:
\begin{align}
\mathbf{p}_0 &= [1,1,1,0,1,0,0,1,1,0,1,1,0,0,0,1], \\
\mathbf{p}_1 &= \overline{\mathbf{p}_0} = [0,0,0,1,0,1,1,0,0,1,0,0,1,1,1,0],
\end{align}
where $\overline{\cdot}$ denotes bitwise negation. Unlike periodic patterns, $\mathbf{p}_0$ is chosen to have strong aperiodic autocorrelation properties, producing a distinct correlation peak that facilitates robust tile localization.

\subsection{Distilled Robust Embedding}
\label{sec:method_distilled_embedding}
While the redundant fountain payload improves recoverability under severe cropping, it inevitably increases the embedding burden. Directly scaling the embedding backbone would lead to higher inference costs and noticeable visual artifacts. To mitigate this, we introduce a training-only distilled embedding branch (\Cref{fig:framework}b and c), transferring robustness-aware behavior from a stronger teacher to a lightweight student pathway.

Specifically, given the stego images $I_{\mathrm{stego}}^{S}$ (student) and $I_{\mathrm{stego}}^{T}$ (teacher), we extract stage-wise features using two Vision Transformer (ViT)~\cite{dosovitskiy2021an} encoders. Distillation is enforced in the feature space: the student features are refined by a Flow-Adapted Residual Attention Alignment (Flow-RAA) module, while the teacher features are transformed by a Flow-Adapted Attention-based Feature Propagation (Flow-AFP) module to serve as supervision targets. This design allows the student to inherit robustness-critical behavior without increasing test-time complexity.

\noindent \textbf{Flow-Adapted Residual Attention Alignment (Flow-RAA).}
At the $l$-th distillation stage, the student ViT outputs a token set $\{F_{l,i}^{S}\}_{i=1}^{N_t}$, where $N_t$ is the number of tokens. As illustrated in \Cref{fig:framework}, Flow-RAA first projects each token, concatenates the projected tokens into a global representation, applies self-attention to model long-range dependencies, and then maps the attended feature back to the token space:
\begin{align}
Z_{l}^{S} &= \operatorname{Concat}\left(\left\{\phi_{\mathrm{proj}}(F_{l,i}^{S})\right\}_{i=1}^{N_t}\right),\\
\hat{Z}_{l}^{S} &= \operatorname{SA}(Z_{l}^{S}),\\
\{\tilde{F}_{l,i}^{S}\}_{i=1}^{N_t}&=\operatorname{Split}\left(\phi_{\mathrm{out}}(\hat{Z}_{l}^{S})\right),
\end{align}
where $\phi_{\mathrm{proj}}(\cdot)$ and $\phi_{\mathrm{out}}(\cdot)$ correspond to the two \textit{Proj} operations in \Cref{fig:framework}, and $\operatorname{SA}(\cdot)$ denotes the self-attention block.

Unlike directly matching token-wise local features, Flow-RAA explicitly aggregates cross-token interactions to capture globally consistent coded structures crucial for crop resilience. The stage-level student descriptor is then obtained via average pooling:
\begin{equation}
\tilde{F}_{l}^{S} = \operatorname{Pool}\left(\{\tilde{F}_{l,i}^{S}\}_{i=1}^{N_t}\right),
\label{eq:raa_pool}
\end{equation}
which serves as the student-side representation for distillation at stage $l$.

\noindent \textbf{Flow-Adapted Attention-based Feature Propagation (Flow-AFP).} Flow-AFP constructs a teacher-side supervision signal that is better aligned with the student representation. Rather than forcing the student to mimic the raw teacher feature directly, Flow-AFP models inter-stage variation in the teacher branch and propagates it onto a student-guided reference.

At the $l$-th stage, Flow-AFP takes as input the propagated student feature from the previous stage $F_{\mathrm{prev}}^{S}$, the teacher feature from the previous stage $F_{\mathrm{prev}}^{T}$, and the current teacher feature $F^{T}$. We first compute the teacher-side feature increment
\begin{equation}
\Delta F_{l}^{T} = F^{T} - F_{\mathrm{prev}}^{T},
\label{eq:afp_increment}
\end{equation}
which corresponds to the subtraction operation in \Cref{fig:framework}. The obtained increment is then fused with $F_{\mathrm{prev}}^{S}$ by a lightweight fusion block:
\begin{equation}
R_{l}^{T} = \psi_{\mathrm{fuse}}\left(F_{\mathrm{prev}}^{S}, \Delta F_{l}^{T}\right),
\label{eq:afp_fuse}
\end{equation}
where $\psi_{\mathrm{fuse}}(\cdot)$ denotes the \textit{Fusion Block} in \Cref{fig:framework}. Finally, the fused residual is added back to the propagated student feature to produce the teacher target:
\begin{equation}
\tilde{F}_{l}^{T} = F_{\mathrm{prev}}^{S} + R_{l}^{T}.
\label{eq:afp_target}
\end{equation}

The subtraction operation effectively suppresses stage-independent bias, isolating newly introduced features, while the residual addition anchors the teacher target to $F_{\mathrm{prev}}^{S}$. This yields a smoothed teacher descriptor $\tilde{F}_{l}^{T}$ perfectly aligned with the student's representational space.

With the outputs of Flow-RAA and Flow-AFP, the distillation loss at the $l$-th stage is defined as
\begin{equation}
\mathcal{L}_{\mathrm{dist}}^{(l)} = \left\| \tilde{F}_{l}^{S} - \tilde{F}_{l}^{T} \right\|_{1},
\label{eq:loss_dist_l}
\end{equation}
and the overall distillation objective is
\begin{equation}
\mathcal{L}_{\mathrm{dist}} = \sum_{l=1}^{N_d} \mathcal{L}_{\mathrm{dist}}^{(l)},
\label{eq:loss_dist_total}
\end{equation}
where $N_d$ is the number of distillation stages. Since the teacher branch is used only during training, the proposed distilled embedding strategy improves the robustness--efficiency trade-off without introducing additional inference overhead to the final student embedding network.

\subsection{Residual Enhancement and Fragment Extraction}
\label{sec:method_enhancer}

The $224\times224$ robust payload image is embedded into the host image using the Attention Affine Coupling Blocks (AACBs)~\cite{ye2025rmsteg} of the AttnFlow model. During extraction, the distorted stego image is passed through the inverse AACBs, yielding a noisy continuous residual prediction $I_{\mathrm{res}}$.

A key difficulty arises from the mismatch between continuous image degradations and discrete $GF(2)$ decoding. While JPEG artifacts, blur, and sensor noise perturb the recovered residual in a continuous domain, the downstream algebraic solver is highly sensitive to bit-flip errors after binarization. To reduce this mismatch, we introduce a lightweight enhancement module, denoted as Enhancer1Ch ($\mathcal{F}_{\mathrm{enh}}$), before the binarization step. Unlike generic image denoisers, $\mathcal{F}_{\mathrm{enh}}$ operates only on the extracted single-channel residual and exploits the strong structural prior induced by the fountain-encoded tile layout. The enhancement process is formulated as

\begin{equation}
    I_{\mathrm{enh}} = \text{clamp}\Big(I_{\mathrm{res}} + \alpha \cdot \tanh\big(\mathcal{F}_{\mathrm{enh}}(I_{\mathrm{res}})\big), \; 0, \; 1\Big)
\label{eq:enhancer}
\end{equation}
where $\alpha$ is a learnable scaling parameter. The $\tanh$ activation helps sharpen ambiguous gray values toward the two binary states.

Trained under progressive cropping-aware perturbations, $\mathcal{F}_{\mathrm{enh}}$ suppresses high-frequency artifacts and restores clearer tile boundaries in the continuous domain. When combined with Otsu's adaptive binarization and pilot-guided filtering, the enhancement module encourages a more decoding-friendly degradation pattern: mild-to-moderate perturbations are corrected before thresholding, while severely corrupted fragments are more likely to be rejected as erasures. In this way, the continuous distortion process is shifted toward an erasure-dominant regime that better matches the assumptions of fountain decoding.

\subsection{GF(2) Decoding}
\label{sec:method_gf2}
\noindent \textbf{Grid Alignment and Extraction.} Given the enhanced residual $I_{\mathrm{enh}}$, we apply Otsu's binarization and perform a 2D sliding window search over all $16 \times 16$ possible phase offsets:
\begin{equation}
(o_x^*, o_y^*) = \arg\max_{(o_x, o_y)} \sum_{t \in \mathcal{T}} \operatorname{match}(t, o_x, o_y, \mathbf{p}_0, \mathbf{p}_1),
\end{equation}
where $\mathcal{T}$ denotes candidate tiles under the current phase hypothesis, and $\operatorname{match}(\cdot)$ measures pilot consistency with the reference patterns $\mathbf{p}_0$ and $\mathbf{p}_1$. The optimal offset dynamically locks onto the remaining tile grid while discarding partial tiles at crop boundaries. Only tiles whose extracted pilot rows exactly match $\mathbf{p}_0$ and $\mathbf{p}_1$ are retained, serving as a pilot-consistency filter for tile acceptance.

\noindent \textbf{Gaussian Elimination over GF(2).} The valid metadata seeds rebuild the pseudo-random bipartite graph, forming a linear system $\mathbf{A} \cdot \mathbf{x} = \mathbf{y}$ over $GF(2)$, where $\mathbf{A} \in \{0,1\}^{m \times K}$ is the coefficient matrix, $\mathbf{x} = [B_0, \dots, B_{K-1}]^T$ denotes the unknown source blocks, and $\mathbf{y}$ contains the observed payloads from the $m$ extracted tiles. We apply Gaussian elimination with XOR operations. If $\text{rank}(\mathbf{A}) = K$, the system is fully determined, and the recovered bitstream is verified against the 32-bit CRC.

\subsection{Recoverability Analysis and Curriculum Motivation}
\label{sec:method_rank_analysis}

By mapping the problem to an erasure channel, the decoding probability becomes closely related to the rank of the recovered linear system. Under a random crop with area retention ratio $r$, the expected number of surviving tiles is $\mathbb{E}[m] = r \cdot N$. Because we enforce four copies of each systematic block among the first $4K = 32$ symbols, under an independence approximation, the probability of losing all copies of any specific systematic block under $r = 0.4$ is reduced to $(1-0.4)^4 \approx 0.129$. The remaining LT-coded symbols then help fill the missing degrees of freedom. Consequently, as long as $m \ge K + \epsilon$, where $\epsilon$ denotes a small decoding overhead, the coefficient matrix tends to reach full rank, which explains the stable recovery behavior observed before the final breakdown threshold.

This rank-based view also motivates our curriculum design. Severe cropping imposes a highly discontinuous optimization signal in early training because small extraction errors may translate into unrecoverable rank deficiency after binarization and decoding. We therefore adopt a progressive cropping-aware training strategy that starts from relatively mild erasure conditions and gradually increases the difficulty, allowing the embedding and enhancement modules to first learn stable localization and residual reconstruction before adapting to harsher cropping regimes.

\begin{table*}[t]
\centering
\caption{Breakdown analysis under coupled regime of mixed distortions and spatial cropping. As the retained image area decreases, CREST degrades more gracefully than competing methods and substantially delays the recovery breakdown boundary. The best results are in \textbf{bold}. See \Cref{sec:experiments_mixed_cropping} for discussion.}
\resizebox{\textwidth}{!}{%
\begin{tabular}{llcccccccccc}
\toprule
\multirow{2}{*}{\textbf{Dataset}} & \multirow{2}{*}{\textbf{Method}} 
& \multicolumn{2}{c}{\textbf{Mixed + $r_{area}$ 1.0}} 
& \multicolumn{2}{c}{\textbf{Mixed + $r_{area}$ 0.7}} 
& \multicolumn{2}{c}{\textbf{Mixed + $r_{area}$ 0.5}} 
& \multicolumn{2}{c}{\textbf{Mixed + $r_{area}$ 0.3}} 
& \multicolumn{2}{c}{\textbf{Mixed + $r_{area}$ 0.15}} \\
\cmidrule(lr){3-4} \cmidrule(lr){5-6} \cmidrule(lr){7-8} \cmidrule(lr){9-10} \cmidrule(lr){11-12}
 & & TRA $\uparrow$ & EMR $\downarrow$ 
   & TRA $\uparrow$ & EMR $\downarrow$ 
   & TRA $\uparrow$ & EMR $\downarrow$ 
   & TRA $\uparrow$ & EMR $\downarrow$ 
   & TRA $\uparrow$ & EMR $\downarrow$ \\
\midrule

\multirow{4}{*}{COCO2017} 
 & DeepMIH~\cite{guan2022deepmih} & 25.1 & 4.514 & 0.00 & 45.21 & 0.00 & 48.33 & 0.00 & 52.14 & 0.00 & 58.60 \\
 & PRIS~\cite{yang2024pris} & 50.1 & 2.6 & 0.00 & 36.54 & 0.00 & 40.12 & 0.00 & 46.88 & 0.00 & 51.35 \\
 & RMSteg~\cite{ye2025rmsteg} & 62.4 & 2.506 & 18.52 & 13.88 & 0.00 & 26.45 & 0.00 & 38.12 & 0.00 & 48.90 \\
 & CREST (Ours) & \textbf{74.2} & \textbf{1.912} 
                  & \textbf{68.45} & \textbf{4.21} 
                  & \textbf{65.12} & \textbf{4.88} 
                  & \textbf{48.55} & \textbf{6.15} 
                  & \textbf{14.20} & \textbf{8.75} \\
\midrule

\multirow{4}{*}{DIV2K} 
 & DeepMIH~\cite{guan2022deepmih} & 21 & 4.049 & 0.00 & 42.15 & 0.00 & 46.50 & 0.00 & 49.30 & 0.00 & 55.45 \\
 & PRIS~\cite{yang2024pris} & 51 & 2.499 & 0.00 & 33.78 & 0.00 & 38.60 & 0.00 & 43.25 & 0.00 & 48.90 \\
 & RMSteg~\cite{ye2025rmsteg} & 78 & 1.65  & 24.36 & 11.45 & 2.15 & 21.60 & 0.00 & 34.50 & 0.00 & 45.20 \\
 & CREST (Ours) & \textbf{85} & \textbf{1.45} 
                  & \textbf{76.50} & \textbf{3.55} 
                  & \textbf{72.30} & \textbf{3.95} 
                  & \textbf{55.40} & \textbf{5.20} 
                  & \textbf{18.65} & \textbf{7.10} \\
\midrule

\multirow{4}{*}{VOC2012} 
 & DeepMIH~\cite{guan2022deepmih} & 24.6 & 4.633 & 0.00 & 46.80 & 0.00 & 50.15 & 0.00 & 54.60 & 0.00 & 61.20 \\
 & PRIS~\cite{yang2024pris} & 50.8 & 2.622 & 0.00 & 37.12 & 0.00 & 42.30 & 0.00 & 48.15 & 0.00 & 53.60 \\
 & RMSteg~\cite{ye2025rmsteg} & 60   & 2.603 & 15.25 & 14.65 & 0.00 & 28.30 & 0.00 & 41.25 & 0.00 & 50.15 \\
 & CREST (Ours) & \textbf{73.5} & \textbf{1.915} 
                  & \textbf{65.80} & \textbf{4.85} 
                  & \textbf{62.45} & \textbf{5.20} 
                  & \textbf{42.10} & \textbf{7.05} 
                  & \textbf{11.35} & \textbf{9.80} \\
\bottomrule
\end{tabular}%
}
\label{tab:mixed_crop}
\end{table*}

\section{Experiments}
\label{sec:experiments}

\subsection{Experimental Settings}
\label{sec:experimental_settings}
We evaluate CREST on three public datasets: COCO2017~\cite{lin2014microsoft}, DIV2K~\cite{agustsson2017ntire}, and VOC2012~\cite{everingham2015pascal}. These datasets cover diverse scene content and image statistics, enabling evaluation across both robustness and visual quality. Following the common setting in robust message embedding, all images are resized to $224 \times 224$.

The comparison includes six baselines: ISN~\cite{lu2021isn}, HiNet~\cite{jing2021hinet}, StegaStamp~\cite{tancik2020stegastamp}, \textbf{DeepMIH}~\cite{guan2022deepmih}, \textbf{PRIS}~\cite{yang2024pris}, and \textbf{RMSteg}~\cite{ye2025rmsteg}. All competing methods are evaluated under aligned distortion settings using their respective test pipelines.

We consider two evaluation regimes. The first is a \emph{coupled distortion-and-cropping} setting, where stego images are first subjected to digital distortions and then evaluated under different area retention ratios. This setting is used to probe the breakdown boundary of message recovery under simultaneous continuous corruption and spatial erasure. The second is a \emph{general distortion} setting, which evaluates robustness, image quality, and deployment efficiency under Gaussian noise, JPEG compression, and mixed distortions.

For image quality, we report \textbf{PSNR}, \textbf{SSIM}~\cite{wang2004image}, and \textbf{LPIPS}~\cite{zhang2018unreasonable}. For decoding robustness, we report \textbf{TRA} (Text Recovery Accuracy) and \textbf{EMR} (Error Module Rate)~\cite{ye2025rmsteg}. In addition, we report \textbf{FPS}, \textbf{GFLOPs}, and \textbf{Params} to evaluate deployment efficiency.

\begin{figure*}[t]
  \centering
  \includegraphics[width=0.8\linewidth]{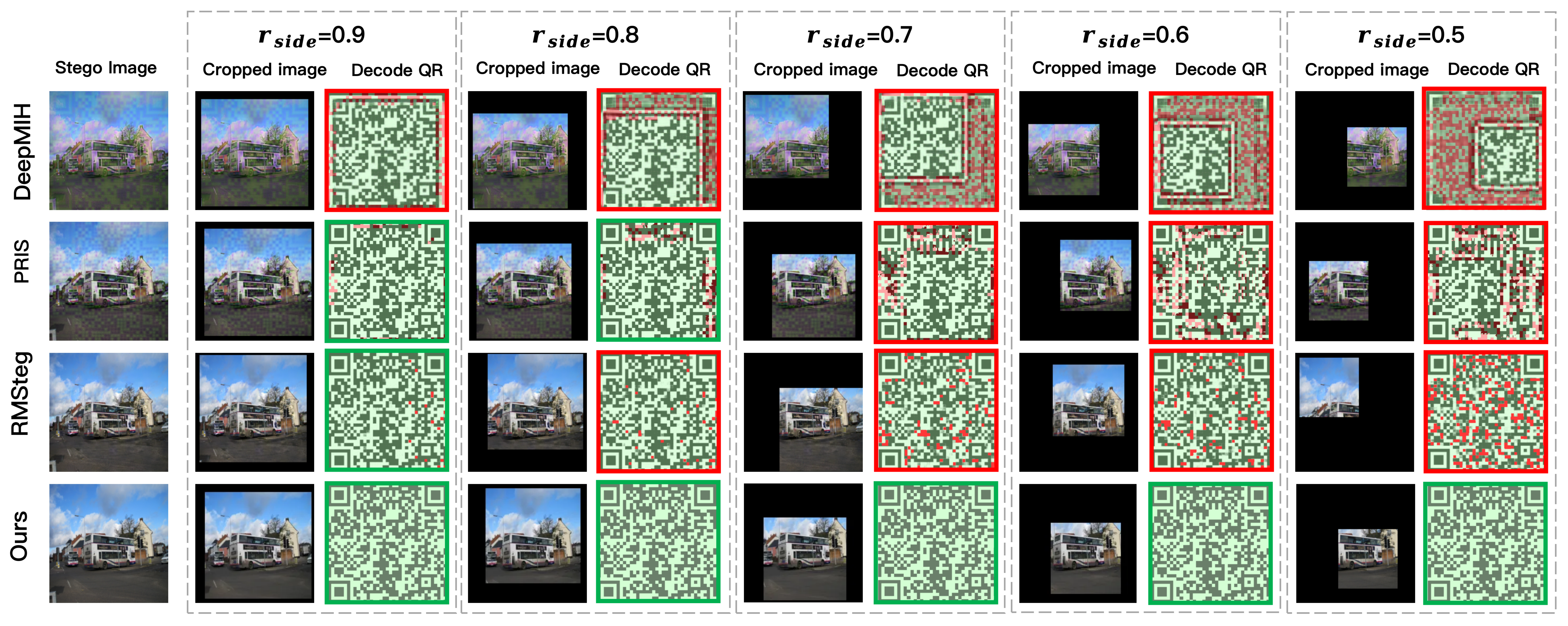}
  \caption{Qualitative decoding results under random pure-cropping attacks with different side retention ratios. Without additional distortions, CREST maintains stable message recovery across different crop locations and crop sizes, indicating that its decoding performance is less sensitive to spatial position under severe cropping. See \Cref{sec:experiments_mixed_cropping} for discussion.}
  \Description{Qualitative results comparing decoded QR codes from DeepMIH, PRIS, RMSteg, and CREST under increasingly severe pure cropping. CREST maintains more complete and decodable QR patterns.}
  \label{fig:pure_crop}
\end{figure*}

\subsection{Coupled Distortions and Cropping}
\label{sec:experiments_mixed_cropping}

\begin{figure*}[t]
  \centering
  \includegraphics[width=0.85\linewidth]{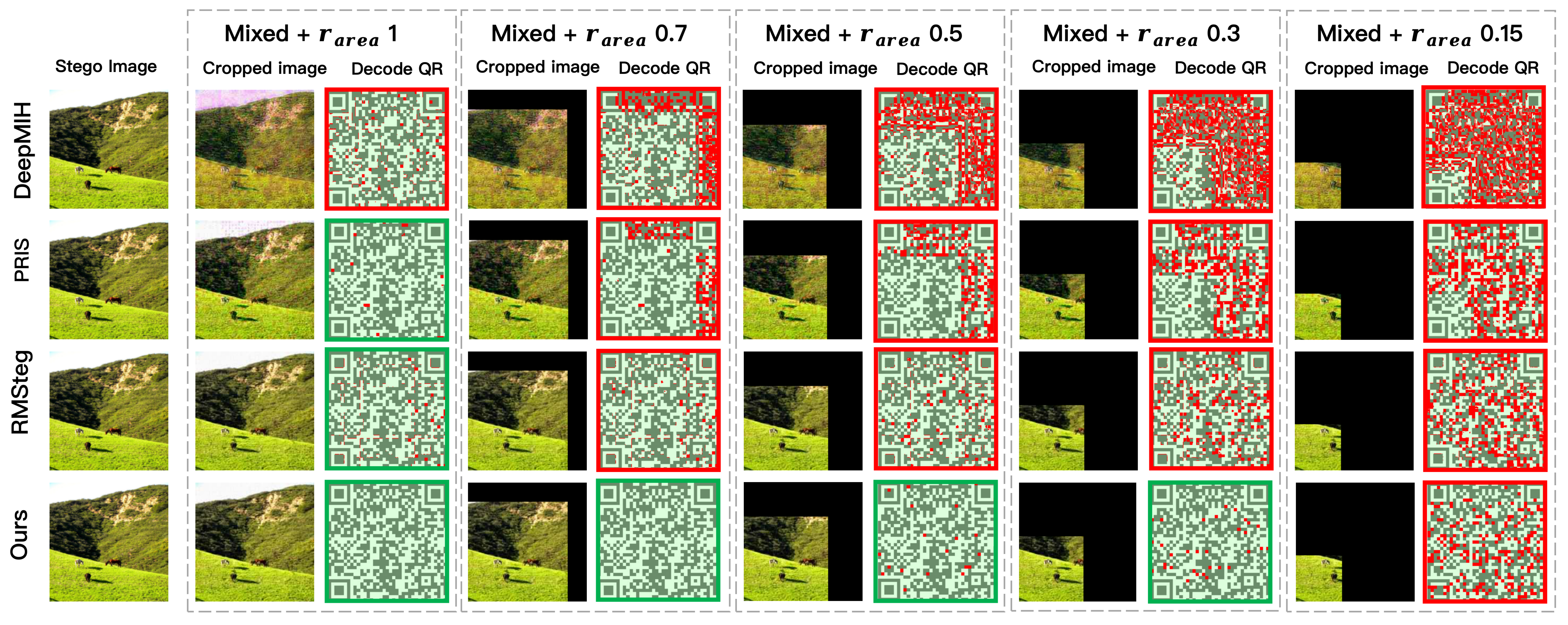}
  \caption{Qualitative comparison under the coupled regime of mixed distortions and spatial cropping. As the retained area decreases, competing methods rapidly lose decodable structure, whereas CREST preserves cleaner decoded QR patterns over a wider range of hostile conditions, indicating a delayed recovery breakdown boundary. See \Cref{sec:experiments_mixed_cropping} for discussion.}
  \Description{A comparison under mixed distortions and spatial cropping. Competing methods rapidly accumulate QR module errors, whereas CREST preserves decodable structure over more severe cropping conditions.}
  \label{fig:mixed_crop}
\end{figure*}

Since CREST achieves near-perfect recovery under pure severe cropping (\Cref{fig:pure_crop}), we focus on the ultimate extreme stress test: a coupled regime of mixed continuous distortions and aggressive spatial erasures ($r_{area}$). As reported in \Cref{tab:mixed_crop} and \Cref{fig:mixed_crop}, CREST consistently dominates across all datasets. Even without cropping ($r_{area}=1.0$), CREST outperforms the strongest baseline, RMSteg~\cite{ye2025rmsteg}. Crucially, this performance gap widens drastically once spatial erasure is introduced. By $r_{area}=0.5$ (discarding half the image), CREST maintains highly robust recovery (e.g., 65.12\% TRA on COCO2017), whereas all competing methods drop to 0.00\% TRA.

This coupled setting exposes a fundamental vulnerability in conventional distortion-robust designs: they suffer catastrophic avalanche failures when spatial support is removed. For instance, DeepMIH and PRIS fail immediately at $r_{area}=0.7$, and RMSteg plummets from 62.4\% to 0.00\% by $r_{area}=0.5$ on COCO2017. In stark contrast, CREST exhibits a graceful, mathematically interpretable breakdown trajectory. As $r_{area}$ is pushed to the extreme limit of $0.15$, the recovery rate steadily declines (from 68.45\% down to 14.20\% on COCO2017) rather than failing abruptly. This confirms the efficacy of our erasure-resilient paradigm: pilot-guided filtering helps convert continuous extraction errors into discrete erasures, enabling the fountain decoder to sustain recovery as long as the surviving valid tiles fulfill the minimum rank requirement.

\subsection{General Distortion, Quality, and Efficiency}
\label{sec:experiments_general_distortion}

\begin{figure}[htbp]
    \centering
    \includegraphics[width=0.99\linewidth]{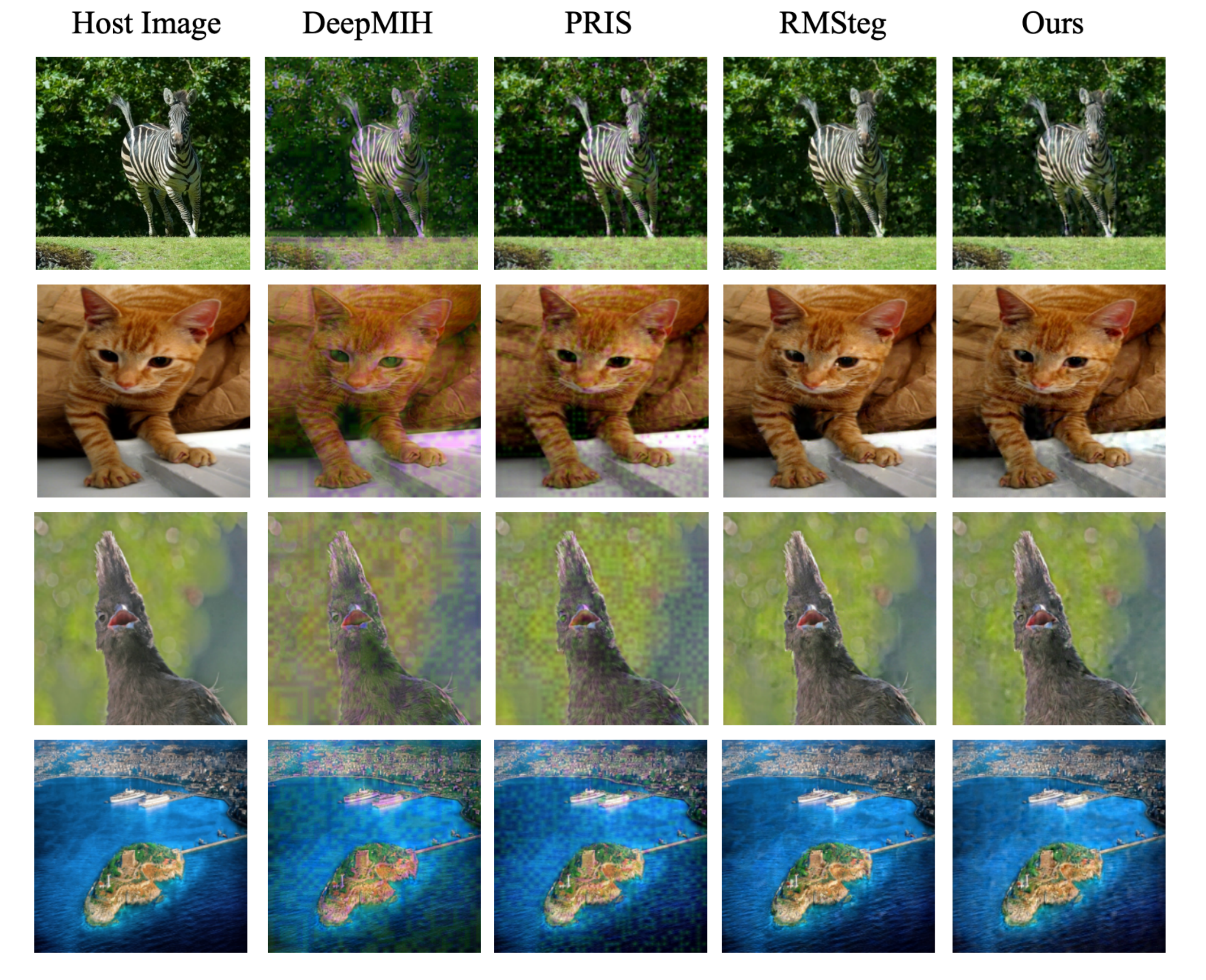}
    \caption{Visual comparison of stego-image quality under mixed distortions. CREST maintains competitive visual quality despite the redundancy introduced for robust recovery. See \Cref{sec:experiments_general_distortion} for discussion.}
    \Description{Visual comparisons of host images and stego images generated by DeepMIH, PRIS, RMSteg, and CREST. CREST largely preserves the visual appearance and semantic content of the host images.}
    \label{fig:mixed_visual}
\end{figure}

Although CREST is primarily designed for severe cropping, it is still important to verify that the proposed erasure-resilient design does not incur prohibitive overhead under conventional distortions. We therefore evaluate image quality, robustness, and deployment efficiency under Gaussian noise, JPEG compression, and mixed distortions. The quantitative and qualitative results are shown in \Cref{tab:mixed,fig:mixed_visual}.

As shown in \Cref{tab:mixed}, CREST remains competitive in image fidelity despite introducing substantial redundancy for recoverability, achieving PSNR/SSIM/LPIPS of 28.854/0.8735/0.1351 under regular conditions. The visual comparisons in \Cref{fig:mixed_visual} further show that CREST preserves clear semantic structures without introducing obvious additional artifacts.

More importantly, CREST achieves the highest FPS and lowest GFLOPs among methods with reported runtime statistics, reaching 47.52 FPS with 74.694 GFLOPs, while also using substantially fewer parameters than RMSteg (114.74 vs.\ 196.79). This indicates that the distilled attention-flow design offsets much of the redundancy cost and yields a favorable robustness--efficiency trade-off.

CREST also remains robust under representative conventional distortions. In particular, it achieves 85.3 TRA under Gaussian noise with $\sigma=0.1$ and 99.5 TRA under JPEG compression with $Q=20$, while maintaining competitive performance under mixed distortions. Overall, these results show that CREST improves severe-cropping robustness without sacrificing practical image quality or deployment efficiency beyond the pure cropping scenario.

\begin{table*}[t]
\centering
\caption{Quantitative comparison under representative general distortions in terms of image quality, robustness, and deployment efficiency. The best result in each column is highlighted in \textbf{bold}. Superscripts indicate the column-wise rank among all methods, with $^{1}$ being the best rank and ties assigned the same rank. TRA and EMR are reported in percentage scale. Although CREST is primarily designed for severe cropping, it remains competitive under conventional distortions while achieving favorable computational efficiency among the compared methods. See \Cref{sec:experiments_general_distortion} for discussion.}
\resizebox{\textwidth}{!}{
\begin{tabular}{l*{16}{c}}
\toprule
\multirow{2}{*}{Method}
& \multicolumn{6}{c}{Regular}
& \multicolumn{2}{c}{$\sigma=0.1$}
& \multicolumn{2}{c}{$\sigma=0.15$}
& \multicolumn{2}{c}{JPEG Q=20}
& \multicolumn{2}{c}{JPEG Q=40}
& \multicolumn{2}{c}{Mixed} \\
\cmidrule(lr){2-7}
\cmidrule(lr){8-9}
\cmidrule(lr){10-11}
\cmidrule(lr){12-13}
\cmidrule(lr){14-15}
\cmidrule(lr){16-17}
& PSNR$\uparrow$ & SSIM$\uparrow$ & LPIPS$\downarrow$ & FPS$\uparrow$ & GFLOPs$\downarrow$ & Params$\downarrow$
& TRA$\uparrow$ & EMR$\downarrow$
& TRA$\uparrow$ & EMR$\downarrow$
& TRA$\uparrow$ & EMR$\downarrow$
& TRA$\uparrow$ & EMR$\downarrow$
& TRA$\uparrow$ & EMR$\downarrow$ \\
\midrule
ISN~\cite{lu2021isn}
& 32.175 & 0.8765 & 0.3266 & - & - & -
& 72.8 & 1.563
& 17.8 & 5.020
& 99.1 & 0.721
& 99.9 & 0.184
& 71.3 & 3.131 \\

HiNet~\cite{jing2021hinet}
& 31.629 & 0.8662 & 0.3423 & - & - & -
& 82.7 & \textbf{1.077}
& 16.2 & 3.724
& 98.6 & 0.573
& 99.7 & 0.099
& 67.7 & 3.426 \\

StegaStamp~\cite{tancik2020stegastamp}
& 21.215 & 0.7027 & 0.3055 & - & - & -
& 5.1 & 6.152
& 0.0 & 10.570
& 95.1 & 1.259
& 97.7 & 0.798
& 55.7 & 3.843 \\

StegaStamp$^{+}$~\cite{tancik2020stegastamp}
& 21.173 & 0.6903 & 0.3418 & - & - & -
& 48.1 & 3.298
& 1.5 & 6.500
& 95.3 & 1.104
& 96.9 & 0.833
& 69.3 & 2.975 \\

DeepMIH~\cite{guan2022deepmih}
& 22.664 & 0.7717 & 0.2993 & 22.59 & 270.517 & 5.40
& 15.6 & 2.856
& 0.0 & 8.727
& 5.1 & 3.443
& 97.0 & 0.773
& 25.1 & 4.514 \\

PRIS~\cite{yang2024pris}
& 28.713 & 0.8573 & 0.1800 & 29.65 & 215.389 & \textbf{4.18}
& 80.4 & 1.818
& 0.0 & 28.500
& 69.3 & 1.742
& 99.0 & 0.208
& 50.1 & 2.600 \\

RMSteg~\cite{ye2025rmsteg}
& \textbf{32.883} & \textbf{0.9109} & \textbf{0.0707} & 30.63 & 107.620 & 196.79
& 79.4 & 1.235
& \textbf{21.6} & \textbf{3.306}
& \textbf{99.5} & 0.117
& \textbf{100.0} & \textbf{0.038}
& \textbf{85.9} & \textbf{0.861} \\

CREST (Ours)
& 28.854$^{4}$ & 0.8735$^{3}$ & 0.1351$^{2}$ & \textbf{47.52}$^{1}$ & \textbf{74.694}$^{1}$ & 114.74$^{3}$
& \textbf{85.3}$^{1}$ & 1.187$^{2}$
& 15.4$^{4}$ & 5.100$^{4}$
& \textbf{99.5}$^{1}$ & \textbf{0.104}$^{1}$
& \textbf{100.0}$^{1}$ & 0.049$^{2}$
& 74.2$^{2}$ & 1.912$^{2}$ \\
\bottomrule
\end{tabular}
}
\label{tab:mixed}
\end{table*}

\subsection{Ablation Study}
\label{sec:experiments_ablation}
We analyze the contributions of three key components in CREST: pilot-guided alignment, spatial anti-clustering, and progressive cropping curriculum learning. The quantitative results are summarized in \Cref{tab:ablation_all}.

\noindent \textbf{Ablation on Aperiodic Pilot Alignment.} As shown in \Cref{tab:ablation_all}(a), removing pilot alignment causes decoding success to collapse across all crop ratios. Even at a mild crop ratio of $r=0.9$, the success rate drops from 100.0\% to 5.5\%. This result confirms that reliable phase locking is essential for stable tile extraction and subsequent $GF(2)$ reconstruction under severe cropping.

\noindent \textbf{Ablation on Spatial Anti-Clustering.} As shown in \Cref{tab:ablation_all}(b), both layouts achieve perfect recovery in the ideal channel, but the non-permuted layout degrades much faster once extraction errors are introduced. For example, at a 15\% extraction error rate, the success rate drops to 55.0\% without permutation, versus 91.0\% with permutation. This shows that global permutation effectively reduces spatial clustering of systematic blocks and mitigates localized rank deficiency.

\noindent \textbf{Ablation on Cropping Curriculum Learning.} As shown in \Cref{tab:ablation_all}(c), training with a fixed severe crop ratio leads to worse optimization behavior, with higher losses and inferior perceptual quality. In contrast, the proposed curriculum achieves lower losses and better visual metrics, indicating a smoother optimization trajectory and a better robustness--imperceptibility trade-off.

Overall, the ablation results support the design logic of CREST: pilot alignment ensures reliable localization, spatial permutation improves fragment diversity under localized erasures, and curriculum learning stabilizes training under progressively harsher cropping conditions.

\begin{table}[t]
\centering
\vspace{-2mm}
\caption{Ablation studies on key mechanisms in CREST. (a) Impact of pilot alignment on decoding success rate; (b) Monte Carlo simulation of spatial anti-clustering under coupled physical stress (15\% edge crop); (c) Comparison of training convergence across different optimization strategies. See \Cref{sec:experiments_ablation} for discussion.}
\vspace{-2mm}
\label{tab:ablation_all}
% Subtable (a)
\vspace{0.1cm}
\resizebox{0.95\columnwidth}{!}{%
\begin{tabular}{l cc c}
\multicolumn{4}{c}{\textbf{(a) Ablation on Aperiodic Pilot Alignment}} \\
\toprule
\textbf{Crop Ratio} & \textbf{w/ Alignment} & \textbf{w/o Alignment} & \textbf{$\Delta$} \\
\midrule
0.9 & \textbf{100.0} & 5.5 & +94.5 \\
0.7 & \textbf{100.0} & 1.5 & +98.5 \\
0.5 & \textbf{100.0} & 0.5 & +99.5 \\
\bottomrule
\end{tabular}%
}

\vspace{0.25cm}
% Subtable (b)
\resizebox{0.95\columnwidth}{!}{%
\begin{tabular}{l cc c}
\multicolumn{4}{c}{\textbf{(b) Ablation on Spatial Anti-Clustering}} \\
\toprule
\textbf{Extraction Error Rate} & \textbf{w/o Perm} & \textbf{w/ Perm} & \textbf{Rank Failure} \\
\midrule
0\% (Ideal Channel) & 100.0\% & 100.0\% & 0  \\
10\% & 78.0\% & 98.5\% & 2200 \\
15\% & 55.0\% & 91.0\% & 4500 \\
20\% & \textbf{31.5\%} & \textbf{83.5\%} & 6850 \\
\bottomrule
\end{tabular}%
}

\vspace{0.25cm}
% Subtable (c)
\resizebox{0.95\columnwidth}{!}{%
\begin{tabular}{l c cc cc}
\multicolumn{6}{c}{\textbf{(c) Ablation on Cropping Curriculum Learning}} \\
\toprule
\textbf{Training Strategy} & \textbf{Epoch} & \textbf{Steg Loss} $\downarrow$ & \textbf{QR Loss} $\downarrow$ & \textbf{SSIM} $\uparrow$ & \textbf{LPIPS} $\downarrow$ \\
\midrule
RMSteg & 5 & 0.0176 & 0.0099 & 0.9211 & 0.0456 \\
Fixed Crop ($r=0.5$) & 5 & 0.0108 & 0.0065 & 0.9608 & 0.0081 \\
\textbf{Ours (Curriculum)} & 5 & 0.0072 & 0.0020 & \textbf{0.9833} & \textbf{0.0031} \\
\midrule
\textbf{Ours (Curriculum)} & 10 & \textbf{0.0065} & \textbf{0.0018} & \textbf{0.9851} & \textbf{0.0024} \\
\bottomrule
\end{tabular}%
}
\end{table}

\section{Conclusion}
In this paper, we revisited robust message embedding under severe local cropping from an erasure-resilience perspective. We argued that, unlike conventional distortions that degrade but preserve the embedded spatial support, severe cropping causes partial payload disappearance and should therefore be addressed as a recoverability problem under missing support. To instantiate this perspective, we proposed CREST, a cropping-resilient and efficient steganographic framework that combines redundancy-enhanced payload construction, cropping-aware robust embedding, and coding-based recovery. Extensive experiments on COCO2017, DIV2K, and VOC2012 demonstrate that CREST substantially improves recovery under severe cropping while maintaining competitive stego-image quality and favorable efficiency. More broadly, our results suggest that robust message embedding under severe cropping may be better approached through the joint design of neural embedding and erasure-resilient decoding, rather than as a straightforward extension of conventional distortion robustness. We believe this paradigm shift paves the way for highly reliable real-world applications, such as physical-to-digital document tracing and tamper-resistant copyright protection. Future work will explore extending this framework to handle more complex physical-world erasures, such as severe perspective warping combined with localized occlusions.

\begin{acks}
This work was supported by the Fundamental Research Funds for the Central Universities at Harbin Engineering University under Grant No. 3072025CFJ0401.
\end{acks}

% \clearpage
\bibliographystyle{ACM-Reference-Format}
\bibliography{samples/4-1}

\end{document}